\documentclass[conference]{IEEEtran}
\IEEEoverridecommandlockouts

\usepackage{cite}
\usepackage{graphicx}
\usepackage{amsmath}
\usepackage{booktabs}
\usepackage{algorithmic}
\usepackage{amssymb}
\usepackage[caption=false,font=normalsize,labelfont=sf,textfont=sf]{subfig}

\begin{document}


\title{Explanation-Guided Federated Deep Reinforcement Learning for Joint Resource Allocation and Scheduling in 6G in-X Subnetworks}
\author{
\IEEEauthorblockN{Ramoni O. Adeogun}
\IEEEauthorblockA{Department of Electronic Systems, Aalborg University, Denmark\\ Email: ra@es.aau.dk}
}

\maketitle

\begin{abstract}
Sixth-generation (6G) wireless systems are envisioned as networks of networks, integrating diverse in-X subnetworks that provide localized, high-performance connectivity. Ensuring reliable communication in dense deployments, such as industrial robots and vehicles, is challenging due to dynamic interference and strict performance requirements. Traditional radio resource management (RRM) methods have limitations, prompting the need for AI-based solutions. In this paper, we address the challenges of dynamic resource allocation and scheduling in 6G in-X subnetworks supporting applications with heterogeneous characteristics by proposing a novel framework that combines Multi-Agent Reinforcement Learning (MARL), Federated Learning (FL), and Explainable AI (XAI). Our solution is designed to improve the reliability, robustness, and transparency of resource management and intra-subnetwork scheduling while ensuring data privacy and fairness across multiple co-existing subnetworks. Unlike existing works, our approach considers a realistic scenario with multiple devices per subnetwork, thereby offering a more comprehensive and scalable solution. The proposed explainable RL framework enables agents to collaboratively optimize channel allocation and scheduling without the need to share raw data, preserving the privacy of each participating subnetwork. Extensive simulations based on 3GPP scenarios demonstrate the effectiveness of our approach, showing significant improvements in performance, and transparency over existing solutions.
\end{abstract}


\IEEEpeerreviewmaketitle

\section{Introduction}
Sixth generation (6G) wireless systems are expected to operate as \emph{networks of networks} integrating subnetworks with heterogeneous types of supported services, varying complexity, coverage, and operating spectrum \cite{berardinelli2023boosting}. 6G in-X subnetworks \cite{adeogun2020towards,berardinelli2021extreme} are short-range, low-power cells envisioned to provide highly localized, high-performance wireless connectivity at the edge of the 6G network of networks. These subnetworks are intended to replace wired infrastructure for communication services with data rate, latency, and reliability requirements that existing wireless technologies cannot support in entities such as robots, production modules, vehicles, and classrooms. Subnetworks are expected to support diverse applications with demanding communication requirements. For example, subnetworks installed within an industrial robot can support high-speed, closed-loop robot control applications with a fraction of a millisecond control cycle time and reliability above $99.9999\%$. Also, classroom subnetworks can enable immersive augmented reality (XR)-based education activities \cite{berardinelli2023boosting}.

The critical nature of these applications supported by 6G subnetworks underscores the importance of ensuring that the required communication requirements are consistently met, regardless of the environment. This becomes especially crucial in dense scenarios, such as subnetworks deployed inside vehicles at a busy intersection, where the potential for high levels of interference could lead to disruption of communication. Consequently, techniques for managing dynamic interference via optimization of radio resource utilization are essential to guarantee the seamless operation of subnetworks in such dense scenarios \cite{adeogun2023distributed}. This has indeed been the focus of active research within the last few years, see e.g., \cite{du2022multi, adeogun2020distributed,adeogun2021learning}. While classical heuristic algorithms for resource management such as distributed greedy selection, centralized graph coloring, minimum SINR guarantee, and sequential iterative subband allocation (SISA) \cite{LiDongSISA} have been studied for subnetworks, there appears to be a consensus that AI-based methods are essential to cope with the requirements for 6G. Since dynamic radio resource management can be modeled as a sequential decision problem, reinforcement learning (RL) based solutions for 6G RRM appear quite popular, see e.g., \cite{khan2023explainable,kavaiya2023learn,6GRRM1}. In the context of 6G in-X subnetworks, the focus has been on Multi-agent RL (MARL) \cite{adeogun2023distributed,du2022multi,adeogun2022multi}. Despite the potential of AI-based resource allocation in existing works, the lack of transparency and trust in the decisions made by an AI model (or RL agent) represents a major bottleneck to the deployment of such solutions in practical systems. Also, privacy and fairness are important to prevent AI models from being discriminatory or biased \cite{marcu2023explainable}. This is particularly relevant for resource allocation in scenarios where multiple independent 6G in-X subnetworks co-exist over a limited set of shared resources. To overcome these challenges, we conjecture that a combination of MARL \cite{du2022multi}, federated learning \cite{roy2023explanation}, and explainable AI (XAI) techniques \cite{ruiz2023xai} can guarantee the reliability, robustness, and transparency of dynamic channel allocation decisions in 6G in-X subnetworks without while preserving the privacy and performance of the system. While MARL provides the framework for solving the dynamic resource allocation problem, FL ensures that MARL agents can be trained collaboratively without explicitly sharing raw measurements between the participating nodes and a central server translating to a guarantee of the privacy of the nodes. On the other hand, XAI provides a set of procedures and methods to reveal the inner operations
of AI algorithms and their potential strengths, weaknesses,
and behaviors. 

This paper addresses three key challenges in dense 6G in-X subnetworks: (i) joint resource allocation and scheduling, (ii) privacy-preserving distributed learning, and (iii) explainable decision-making. The main contributions are as follows: 
\begin{itemize}
    \item We propose a novel AI-based solution for joint resource allocation and scheduling in dense deployments of 6G in-X subnetworks based on a combination of MARL, FL, and XAI. To the best of our knowledge, this is the first work on joint resource allocation and scheduling in the context of 6G in-X subnetworks. Unlike existing works on RRM for subnetworks, e.g., \cite{adeogun2020distributed, adeogun2022multi} which make the unrealistic assumption of a single device per subnetwork, this paper considers a more general and realistic set up with multiple devices in each subnetwork. 
    \item We design an explainable RL framework for training agents to perform joint channel allocation and intra-subnetwork scheduling.
    \item We perform extensive simulations using 3GPP channel models and benchmark performance of the proposed scheme with state-of-the-art solutions.  
\end{itemize}

\section{System Model and Problem Formulation}
\subsection{System Model}
In this paper, we consider the deployment of $N$ 6G in-X subnetworks in an indoor factory hall for supporting industrial control operations. The subnetworks are indexed by the set $\mathcal{N}=\{1,2,\cdots, N\}$.  Each subnetwork consists of a controller collocated with a single access point (AP) for coordinating transmission to/from its associated devices i.e., sensor and actuators). We assume the $n$th subnetwork supports $M_n$ randomly distributed devices. The devices in the $n$th subnetwork are indexed with $m_n\in \mathcal{M}_n = \{1,2,\cdots, M_n\}$. We assume that a total bandwidth, $B_{s}$, which is partitioned into $Z$ sub-bands is available and that the number of sub-bands is much less than the number of subnetworks, i.e., $Z<<N$. We index the available sub-bands with $z\in \{1,2,\cdots, Z\}$. Denoting the transmit power as $P$, the received signal strength (RSS) on the link between the $n$th AP from the $m$th device in the $u$th subnetwork can be expressed as
 \begin{equation}
     \label{eq:rxPow}
     \xi^{k}_{n,u,m}[t] = P|h^{k}_{n,u,m}[t]|^2\Gamma^k_{n,u,m}\vartheta_{n,u,m},
 \end{equation}
 where $\Gamma^k_{n,u,m}$, $h^{k}_{n,u,m}[t]$, and $\vartheta_{n,u,m}$ denote the pathloss,  the small scale gain and log-normal shadowing, respectively. The small scale gain, $h^{k}_{n,u,m}[t]$, is modelled as
 \begin{equation}
     h^{k}_{n,u,m}[t] = \rho h^{k}_{n,u,m}[t-1]+\sqrt{1-\rho^2}\varrho^k_{n,u,m},
 \end{equation}
where $\rho$ is the  autocorrelation coefficient and $\varrho^k_{n,u,m}$ is an iid complex Gaussian variable. The autocorrelation coefficient is modeled as \[\rho = J_0(2\pi f_dT_s)\] 
where $J_0(\cdot)$, $T_s$ and $f_d$ denote the zeroth order Bessel function of the first kind, the slot duration, and the maximum Doppler frequency, 
respectively. 

The path-loss component, $\Gamma^k_{n,u,m}$ is expressed as $\Gamma^k_{n,z,m} = c^2d_{n,u,m}^{-\alpha}/16\pi^2f_k^2$,
    where $d_{n,u,m}$ is the link distance, $c\approx 3\times 10^8~\text{ms}^{-1}$ denotes the speed of light, $f_k$ and $\alpha$ are the carrier frequency of channel $k$ and path-loss exponent, respectively. The shadowing fading component is computed using
    \begin{equation}
        \label{eq:eqSF}
        \vartheta_{n,u,m} = \ln\left\{\frac{1-e^{\left(-\frac{d_{n,u,m}}{d_{corr}}\right)}}{\sqrt{2}\sqrt{1+e^{\left(-\frac{d_{n,u,m}}{d_{corr}}\right)}}}\left(\mathbf{Y}_n+\mathbf{Y}_{u,m}\right)\right\},
    \end{equation}
    where $\mathbf{Y}_x$ is the value of a two-dimensional Gaussian random field at the device's or AP's location, and $d_{corr}$ is the correlation distance.

We assume that transmission within each subnetwork occurs over a single sub-band and that each sub-band is further divided into $L (L\leq M_n;  \forall n)$ resource units (RUs). Devices within each subnetwork are then scheduled over the $L$ RUs. With these assumptions, each transmission with a subnetwork may experience both intra- and inter-subnetwork interference. At slot, $t$, the received signal-to-noise-plus-interference ratio (SINR) on the link between the $n$th AP and its $m$th device can therefore be expressed as
\begin{equation}
\label{SINR1} 
    \Upsilon_{nm}^{z}[t] = \frac{\xi^k_{n,n,m}[t]}{\sum_{n'\in\mathcal{J}_{nn'}}\xi^z_{n,n',m'}[t]+\sum_{n'\in\mathcal{I}_{nn'}}\xi^z_{n,n',m'}[t] + \sigma^2}
\end{equation}
where $\mathcal{J}_{nn'}$ and $\mathcal{I}_{nn'}$ denotes the set of all devices and AP generating intra- and inter-subnetwork interference on the $z$ subband, respectively. The term, $\sigma^2$ denotes the noise power calculated as a function of the bandwidth of each RU. Assuming a single antenna at both the APs and devices and considering the finite-block length approximation, the achieved rate can then be expressed as 
\begin{equation}
    \zeta_{nm}[t] \approx \log_2(1+\Upsilon_{nm}[t]) - \sqrt{\frac{V_{nm}[t]}{C_{\ell}}}Q^{-1}(\eta)\log e,
\end{equation}
where $\eta$ denotes the decoding error probability, $Q$ is the complementary Gaussian cumulative distribution function, and $V$ denotes the channel dispersion which is defined as
\begin{equation}
    V_{nm}[t] = 1-\frac{1}{(1+\Upsilon_{nm}[t])^2}.
\end{equation}
\subsection{Control Operation Characteristics}
\label{sec:Xtics}
Unlike recent works on resource allocation for 6G in-X subnetworks, we consider more realistic heterogeneous characteristics of the control operations within each subnetwork and the associated communication characteristics. We consider the following key assumptions about the control operations supported by the devices in each subnetwork: 
\begin{itemize}
    \item \textbf{Packet size: }The packet size of sensors is determined by their function. For instance, while sensors such as light, temperature, contact, proximity, and ultrasonic typically require a small packet size (in the order of a few bytes), others such as image sensors, audio sensors, and environmental sensors like air quality sensors require a larger packet size (in the order of kilobytes to megabytes). We, therefore, consider these two classes of devices within each subnetwork. The packet size for device $m$ transmission in subnetwork, $n$ is then denoted as
    \begin{equation}
        \label{eqPacket}
        \beta_{mn} = \beta_{\mathrm{small}}\cdot \mathbb{I}(U \leq p) + \beta_{\mathrm{large}} \cdot \mathbb{I}(U > p),  
    \end{equation}
    where $U\sim\mathcal{U}(0,1)$ is a uniform random variable, $p (0\leq p\leq 1$ is the probability that a device has packet size, $\beta_{\mathrm{small}}$ and $\mathbb{I}$ is the indicator function that equals $1$ if the condition is true and $0$ otherwise. 
    \item \textbf{Packet periodicity: } Similar to the works in \cite{adeogun2020distributed, adeogun2022multi}, we consider deterministic periodic traffic with different periodicity depending on the requirements of the control operation supported by the device. We denote the period for the $m$th device in subnetwork $n$ as $T_{mn}$.
    \item \textbf{Survival time: } Another important characteristic of control applications is the survival time defined as the number of slots over which a control operation can continue without successful reception of anticipated packets. The survival time of the control operation supported by device $m$ in the $n$th subnetwork is denoted as $\tau_{mn}$. 
\end{itemize}

With the above control operation characteristics, intra- and inter-subnetwork interference becomes inevitable and must be properly managed to guarantee stringent communication requirements. 
\subsection{Problem Formulation}
We consider an RRM problem involving a fully distributed joint selection of sub-bands and intra-subnetwork scheduling. Considering the control characteristics in Section~\ref{sec:Xtics}, the optimization problem can be formulated as that of minimizing the probability that the control operation fails or equivalently the probability that the burst error length, $L_{\mathrm{burst}}$ (i.e., the number of consecutive transmission opportunities in which a device fails to successfully deliver its packet) exceeds the survival time for all devices and formally be written as 
\begin{align}\label{eq:eqProb}
    \text{P}: & \left\{\left\{\min_{\{\mathbf{a},\mathbf{v}\}} \operatorname{Prob}[L_{\mathrm{burst}}(\mathbf{a},\mathbf{v}_n)^{mn}>\tau_{mn}]\right\}_{m=1}^{M_n}\right\}_{n=1}^N\nonumber\\
    \text{s.t.} & \quad |\mathbf{v_n}| \leq L \quad \forall n
\end{align}
where $\mathbf{a} = [a_1\cdots a_N]; \, a_n \in \{1,2,\cdots, Z\}\, \forall n$ is a vector of indices of the sub-band selected by all subnetworks and $\mathbf{v}$ contains the device scheduling decisions of all subnetworks. Note that minimizing the probability of control-operation failure as in \eqref{eq:eqProb} directly improves communication reliability for time-sensitive industrial applications.

\begin{figure*}[t]
\vspace{10pt}
    \centering
    \includegraphics[scale=0.62]{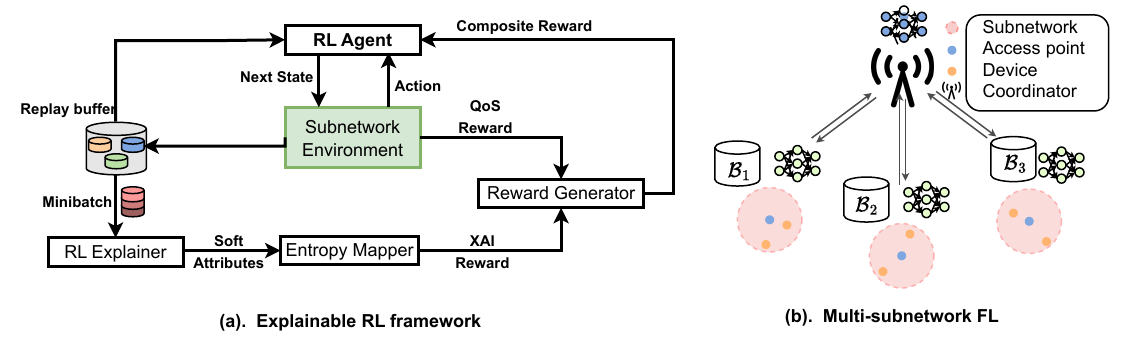}
    \caption{Illustration of the framework for explainable reinforcement learning (a) and multi-subnetwork federated learning (b).}
    \label{fig:EFDRLFL}\vspace{-10pt}
\end{figure*}

\section{Proposed Explainable DRL Method for Resource Allocation}
To solve the multi-objective optimization problem in \eqref{eq:eqProb}, we proposed an explainable federated DRL (EFDRL) for joint sub-band selection and intra-subnetwork scheduling in this paper. The suggested EFDRL framework is depicted along with an illustration of the federated training procedure in Figure~\ref{fig:EFDRLFL}. EFDRL combines explanation-guided learning (EGL) with federated multi-agent DRL to obtain enhanced performance and interoperability of the agent's decisions.  The different components of the proposed approach are described in the sequel. 
\subsection{Action Space}
The decisions to be taken by the agents involve two main components viz: sub-band selection and scheduling to minimize inter- and intra-subnetwork interference, respectively. The action space for the $n$th subnetwork is then formed by combining the set of all available sub-bands and the set containing all possible combinations of $L$ out of its $M_n$ devices.   

\subsection{State Space}
To aid the agents in learning to take joint sub-band selection and intra-subnetwork scheduling decisions, the local observation for each subnetwork is defined as a combination of the current wireless link characteristics (i.e., channel gain between the AP and all its associated devices), $\{\xi_{n,n,m}(t)\}_{m=1}^{M_n}$, the control system status (current estimated time to failure), $\{\tau_{nm}^{\mathrm{TTF}}(t)\}_{m=1}^{M_n}$ \footnote{Time-to-Failure (TTF) represents the remaining number of transmission opportunities before the survival time threshold of device mmm in subnetwork nnn is violated.} and the sense aggregate interference power on all sub-bands, $\{\mathcal{I}_n^z(t)\}_{z=1}^{Z}$. In addition, the current action, $a_n(t-1)$ and reward value are included as part of the local state.  The local observation for subnetwork $n$ is then defined as
\begin{align}
\label{eq:statelocal}
\mathbf{s}_n[t] &= \left[\{\xi_{n,n,m}(t)\}_{m=1}^{M_n}\,\, , \{\tau_{nm}^{\mathrm{TTF}}(t)\}_{m=1}^{M_n}\,\, ,\{\mathcal{I}_n^z(t)\}_{z=1}^{Z},\right.\nonumber\\
&\quad\quad\quad\quad\left. a_n(t-1), r_n(t-1)\right]^T
\end{align}

\subsection{Explanation Guided Reward Signal}
Inspired by the works in \cite{rezazadeh2023explanation, roy2023explanation}, we design a composite reward comprising a QoS and XAI denoted for the $n$ subnetwork as $r_n^{\mathrm{QoS}}$ and $r_n^{\mathrm{XAI}}$, respectively. The composite reward, which is used as feedback during the training, is given as 
\begin{equation}
    r_n(t) = r_n^{\mathrm{QoS}}(t)+r_n^{\mathrm{XAI}}(t).
\end{equation}
Considering the optimization problem in \eqref{eq:eqProb}, the QoS reward for the subnetwork $n$ is designed as
\begin{equation}
    r_n(t) = \frac{1}{M_n}\sum_{i=1}^{M_n} r_{nm}(t),
\end{equation}
where $r_{nm}(t)$ denotes the reward received by the agent in the $n$th subnetwork due to the success or otherwise of the link between sensor $m$ and the AP and is defined as
\begin{equation}
\label{eq:reSensor}
  r_{n}(t) =
    \begin{cases}
      +\alpha_{\mathrm{s}} & \text{if transmission is succesful}\\
      \frac{-\alpha_{\mathrm{f}}}{\exp{(\beta\mathrm{TTF}_n(t))}} & \text{otherwise}
    \end{cases}.       
\end{equation}
In \eqref{eq:reSensor}, $\alpha_{\mathrm{s/f}} (\alpha_{\mathrm{s/f}} > 0)$ is a constant reward/penalty scaling factor, and $\beta (0<\beta\leq 1)$ is a parameter that controls the decay of the exponential function depending on the time to failure, $\mathrm{TTF}_{nm}(t)$ of the control operation supported by the sensor $m$. 

To characterize the XAI component of the reward, we adopt the multiplicative inverse of the Shannon entropy, i.e., 
\begin{equation}
    r_n^{\mathrm{XAI}}(t) = \frac{1}{\max_{u}\mathcal{H}_u},
\end{equation}
with the entropy, $\mathcal{H}_u$ expressed as 
\begin{equation}
\label{eq:entropy}
    \mathcal{H}_u = -\sum_{z=1}^Z p_{z,u}\log(p_{z,u}).
\end{equation}
In \eqref{eq:entropy}, $p_{z,u}$ denotes the probability distribution of the states-features and is expressed for the $u$th sample in the experience replay buffer of the agent in subnetwork $n$ as
\begin{equation}
    p_{n,z,u}= \frac{\exp\{|\eta_{n,z,u}|\}}{\sum_{z'=1}^Z \exp\{|\eta_{n,z',u}|\}},
\end{equation}
where $\eta_{n,z,u}$ denotes the SHAP value computed for state variable $z$ of sample $u$ in the experience replay buffer. 
\subsection{Policy Representation}

To model the mapping between the state measurements and the dynamic joint channel selection and scheduling decision at each subnetwork, we propose to use a multi-agent proximal policy optimization (MAPPO) \cite{yu2022surprising} algorithm with federated training. We denote the proposed algorithm as F-MAPPO. In F-MAPPO, two separate networks are trained for each subnetwork: an actor network and a value function network (called a critic). Without loss of generality, we assume that all agents share the actor and critic network. We denote the actor and critic network as $\pi_\theta$  and $V_\phi$, respectively. The actor-network learns to map the agent observations to a categorical distribution over the actions and is trained to maximize the objective function \cite{yu2022surprising}
\begin{align}
    \label{eq:actorLoss}
    \mathcal{L}(\theta) &= \frac{1}{|\mathcal{B}_n|N}\sum_{b=1}^{|B|}\sum_{n=1}^N\min \left(r_{\theta,b}^{n}A_{b}^{n},\operatorname{clip}\left(r_{\theta,b}^{n},1-\epsilon,1+\epsilon\right)A_{b}^{n}\right)\nonumber\\
    &\quad\quad +\frac{\sigma}{|\mathcal{B}_n|N}\sum_{b=1}^{|B|}\sum_{n=1}^N\mathcal{S}\left[\pi_\theta(s_b^n)\right],
\end{align}
where $|\mathcal{B}|$ and $\sigma$ denote the batch size and the entropy coefficient, respectively. The term $\mathcal{S}$ represents the policy entropy, $A_b^n$ denotes the advantage, which is computed using a generalized advantage estimation \cite{GAEpaper} method, and $r_{\theta,b}^{n}$ is defined as
\begin{equation}
    r_{\theta,b}^{n} = \frac{\pi_\theta(a_b^n|s_b^n)}{\pi_{\theta_{old}}(\pi_\theta(a_b^n|s_b^n)}.
\end{equation}
The critic network learns to minimize the function
\begin{align}
    \label{criticloss}
    \mathcal{L}(\phi) &= \frac{1}{|\mathcal{B}_n|N}\sum_{b=1}^{|B|}\sum_{n=1}^N \max \left[ \left(V_\phi(s_b^n)-\hat{R}_b\right)^2, \right.\nonumber\\
    & \left.\left(\operatorname{clip}\left(V_\phi(s_b^n), V_{\phi_{old}}(s_b^n)-\epsilon,V_{\phi_{old}}(s_b^n)+\epsilon\right)-\hat{R}_b\right)^2\right],
\end{align}
where $\hat{R}_b$ denotes the discounted reward calculated over data $b$ from the experience replay buffer.    

\begin{table}[!]
\centering
\caption{Simulation Parameters}
\label{tab:simulation_parameters}
\scalebox{0.85}{
\begin{tabular}{l|c}
\toprule
\textbf{Parameter}             & \textbf{Value}                    \\ \midrule
Number of subnetworks          & 10                                \\ 
Devices per subnetwork         & 4                                 \\ 
Number of subchannels          & 3                                 \\ 
Number of resource units       & 2                                 \\ 
Subnetwork radius              & 0.5 m                             \\ 
Minimum device-controller distance & 0.3 m                         \\ 
Minimum controller distance    & 1.0 m                             \\ 
Channel bandwidth              & 40 MHz                         \\ 
Carrier frequency              & 6 GHz                             \\ 
Clutter type                   & Dense                             \\ 
Clutter element size           & 2 m                               \\ 
Clutter density                & 60\%                              \\ 
Shadowing standard deviation   & 7.2                               \\ 
Maximum transmission power     & 1 W                               \\ 
Noise power spectral density   & -174 dBm/Hz                       \\ 
Noise figure                   & 5 dB                              \\ 
Factory area                   & $20 \times 20$ m$^2$              \\ 
Correlation distance           & 5 m                               \\ 
Sample time                    & 0.05 s                            \\ 
Robot speed                    & 5 m/s                             \\ 
Target transmission rate       & Randomized (0.4 or 1 Mbps)        \\ 
Survival times                 & Randomized (2-5 steps)            \\ 
Traffic periodicity            & Randomized (1-10 steps)           \\ \bottomrule
\end{tabular}
}
\end{table}
\begin{figure}
    \centering
    \includegraphics[width=0.65\linewidth]{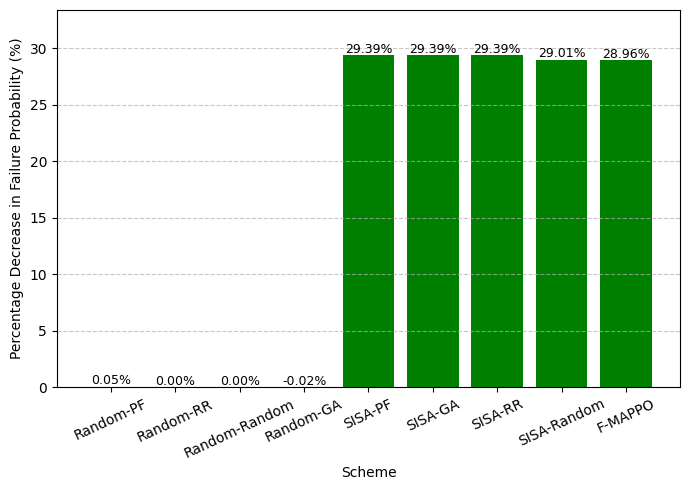}
    \caption{Decrease in probability of failure relative to random subband allocation with random scheduling with $M = 2$. }
    \label{fig:fig2}\vspace{-10pt}
\end{figure}
\begin{figure}
    \centering
    \includegraphics[width=0.65\linewidth]{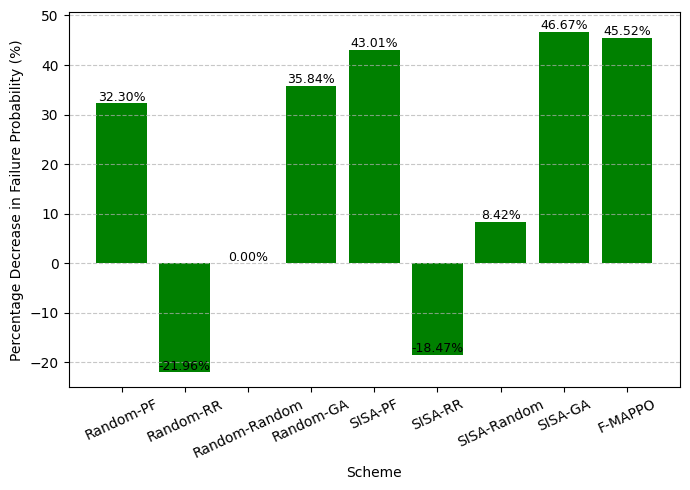}
    \caption{Decrease in probability of failure relative to random subband allocation with random scheduling with $M = 4$.}
    \label{fig:fig3}\vspace{-10pt}
\end{figure}
\begin{figure}[ht]
    \centering
    \includegraphics[width=0.65\linewidth]{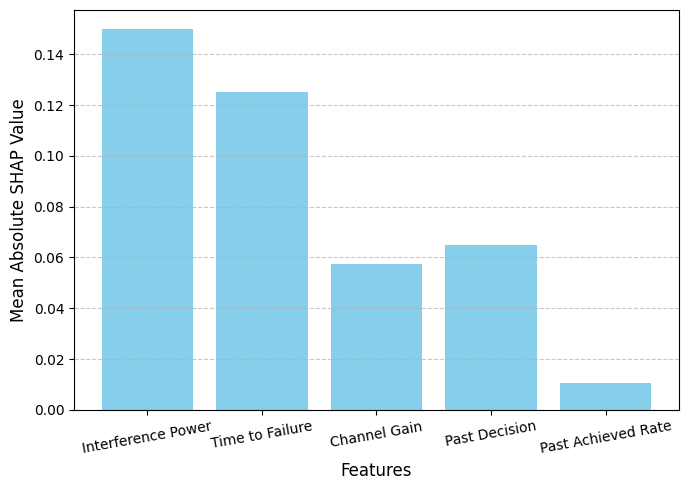}
    \caption{Mean absolute SHAP values for the features in the state space.}
    \label{fig:fig4}\vspace{-12pt}
\end{figure}
\section{Performance Evaluation}
\subsection{Simulation Setup}
We consider a network with $N = 10$ subnetworks, each comprising a single controller serving as the access point (AP) for multiple devices within the subnetwork. The subnetworks are uniformly distributed within a rectangular area $20 \, \text{m} \times 20 \, \text{m}$, corresponding to a high deployment density of $25,000 \, \text{subnetworks/km}^2$. Subnetworks move in the area following a restricted random direction mobility with a velocity $v = 2 \, \text{m/s}$, resulting in a Doppler frequency of $f_d = 40 \, \text{Hz}$ at the considered carrier frequency of $6~\text{GHz}$. A total bandwidth $B = 120 \, \text{MHz}$, which is divided into $K = 3$ subbands, each consisting of $Z=2$ resource units (RUs) is considered. We utilized a fixed transmit power $P_\text{tx} = 0 \, \text{dBm}$. In the simulation, each subband is divided into 2 resource units (RUs), enabling the concurrent transmission of up to $2$ devices within each subnetwork at every time instant. At each transmission slot, $Z_s (1\leq Z_s\leq Z)$ devices are scheduled on the RUs within the subband allocated to each subnetwork. Devices are associated with varying target rates, randomly assigned from a low-rate and high-rate distribution, reflecting the diverse communication demands of industrial applications. Each device has a survival time randomly selected between $2$ and $6$ slots, representing critical deadlines for successful transmission. Other simulation parameters are presented in Table~\ref{tab:simulation_parameters}.

\subsection{Benchmarks}

The performance of the proposed method is compared to the following state-of-the-art benchmarks:
\begin{itemize}
    \item \textbf{SISA - PF}: This approach combine the Subband Iterative Scheduling Algorithm (SISA) with Proportional Fair scheduling. This approach balances fairness and efficiency by leveraging SISA for subband allocation while optimizing scheduling to prioritize users with favorable channel conditions relative to their time to failure.
\item \textbf{SISA - RR}: This method uses SISA for subband allocation but employs a Round Robin (RR) scheduler approach to ensure fairness among users. Each device is served in turn, irrespective of channel conditions or requirements.

\item \textbf{Random - PF}: Subbands are randomly allocated to subnetworks. Proportional Fair scheduling is then applied to perform intra-subnetwork scheduling.

\item \textbf{Random - RR}: Subbands are assigned randomly, and devices are scheduled via round robin.

\item \textbf{Random - Random}: Both subband allocation and scheduling are completely random. This serves as a baseline to assess the performance of other algorithms.

\item \textbf{SISA - GA}: This method combines SISA with an idealized scheduler that has perfect knowledge of all network parameters. This provides an upper bound on performance.
\end{itemize}

\subsection{Results}

The F-MAPPO-based approach for joint subband allocation and intra-subnetwork scheduling is trained over $3,000$ episodes, each consisting of $200$ steps with an interval of $0.05~\text{s}$ between steps. During training, the agent demonstrates convergence after approximately $2,000$ episodes. The trained agent is then deployed across each subnetwork to perform joint subband selection and scheduling during the execution phase, which spans $500$ episodes. The performance of the proposed scheme is evaluated based on the probability of failure, defined as the likelihood that the burst error length at any device exceeds the survival time. To quantify the improvement resulting from a subband allocation and/or scheduling approach, we compute the percentage decrease of the probability of failure (PDPF) of the different schemes relative to the Random - Random baseline which performs both subband allocation and scheduling randomly.  

Figure~\ref{fig:fig2} presents the PDPF for various approaches when the number of devices equals the number of RUs per unit, i.e., $M = Z = 2$. As expected, all schemes based on random subband allocation show no improvement in PDPF. This outcome is due to the scheduling decision, which permits the transmission of all devices, as the number of available RUs matches the number of devices. Conversely, methods utilizing SISA for subband allocation achieve a significant improvement in the probability of failure, approximately $29\%$. The proposed F-MAPPO method offers a comparable gain of $28.96\%$, resulting in similar performance but with substantially lower complexity.

The PDPF is shown in Figure~\ref{fig:fig3} for the case where the number of devices per subnetwork is $M = 4$ and the number of RUs is $Z = 2$, corresponding to a scenario where both subband allocation and scheduling decisions impact overall performance. The figure demonstrates that using a round-robin (RR) scheduler is detrimental to transmission performance, regardless of the scheme used for subband allocation. The performance degradation due to RR scheduling is approximately $22\%$ and $18\%$ with random and SISA-based subband allocation, respectively. This is expected, as RR scheduling ignores the survival time requirement, potentially delaying transmissions that are critical for devices with tighter deadlines. The figure also reveals that survival time-aware PF scheduling and Genuine Aided (GA) schedulers result in significant performance improvements, even when random subband allocation is used. Overall, the SISA-GA and F-MAPPO schemes offer the best performance, with a reduction in the probability of device transmission failure of $46.67\%$ and $45.52\%$, respectively.

Figure~\ref{fig:fig4} presents a bar chart of the mean absolute SHAP values, highlighting the relative importance of various features in the state space within the model's decision-making process for joint scheduling and subband allocation. As illustrated, features such as interference power and time to failure exhibit the highest SHAP values, underscoring their critical role in shaping the agent's decisions. This aligns with domain knowledge, as mutual interference between subnetworks directly affects the efficiency of subband allocation, while time-to-failure is essential for prioritizing devices appropriately in scheduling. The figure also reveals that intra-subnetwork channel gain and past decisions have relatively lower impact compared to interference power and time to failure. This suggests that the model uses these features to refine its predictions, but with less emphasis than the more influential features. In contrast, features like past achieved rate show comparatively lower SHAP values, indicating a minimal direct influence on decisions, although they may contribute indirectly through interactions with other features.

\section{Conclusion}

In this paper, we proposed a novel AI-based framework for dynamic resource allocation and scheduling in dense deployments of 6G in-X subnetworks. By integrating Multi-Agent Reinforcement Learning (MARL), Federated Learning (FL), and Explainable AI (XAI), our approach effectively addresses key challenges such as privacy, transparency, and scalability. Unlike existing solutions, our method supports multiple devices per subnetwork, providing a more realistic and flexible resource management strategy that can adapt to the complexities of 6G networks. The proposed framework enables collaborative optimization without the need for raw data sharing, ensuring the privacy of individual subnetworks. Additionally, the use of XAI techniques enhances the interpretability of the decision-making process, allowing stakeholders to better understand and trust the model’s decisions. Our extensive simulations, based on 3GPP scenarios, demonstrate that the proposed approach outperforms existing baselines for joint subband allocation and scheduling, which combine state-of-the-art solutions for subband allocation and scheduling. The results confirm that our framework not only improves performance but also provides a scalable and transparent solution for managing resource allocation.

\bibliographystyle{IEEEtran}
\bibliography{bibtex/bib/XAISubnetworkbib}
\end{document}